\documentclass[pdflatex,sn-mathphys-num]{sn-jnl}
\usepackage{algorithm}
\usepackage{algpseudocode}
\usepackage{graphicx}%
\graphicspath{ {./figures/} } 
\usepackage{multirow}%
\usepackage{amsmath,amssymb,amsfonts}%
\usepackage{amsthm}%
\usepackage{mathrsfs}%
\usepackage[title]{appendix}%
\usepackage{xcolor}%
\usepackage{textcomp}%
\usepackage{manyfoot}%
\usepackage{booktabs}%
\usepackage{algorithmicx}%
\usepackage{algpseudocode}
\usepackage{listings}%
\theoremstyle{thmstyleone}%
\theoremstyle{thmstyletwo}%
\theoremstyle{thmstylethree}%
\begin{document}

\title[Article Title]{Retrieval Augmented Generation Based LLM Evaluation For Protocol State Machine Inference With Chain-of-Thought Reasoning}

\author[]{Youssef Maklad, Fares Wael, Wael Elsersy, and Ali Hamdi}
\affil[]{Faculty of Computer Science, MSA University, Giza, Egypt}
\affil[]{\texttt{\{youssef.mohamed88, fares.wael, wfarouk, ahamdi\}@msa.edu.eg}}

\abstract{This paper presents a novel approach to evaluate the efficiency of a RAG-based agentic Large Language Model (LLM) architecture for network packet seed generation and enrichment. Enhanced by chain-of-thought (COT) prompting techniques, the proposed approach focuses on the improvement of the seeds' structural quality in order to guide protocol fuzzing frameworks through a wide exploration of the protocol state space. Our method leverages RAG and text embeddings to dynamically reference to the Request For Comments (RFC) documents knowledge base for answering queries regarding the protocol's Finite State Machine (FSM), then iteratively reasons through the retrieved knowledge, for output refinement and proper seed placement. We then evaluate the response structure quality of the agent's output, based on metrics as BLEU, ROUGE, and Word Error Rate (WER) by comparing the generated packets against the ground-truth packets. Our experiments demonstrate significant improvements of up to 18.19\%, 14.81\%, and 23.45\% in BLEU, ROUGE, and WER, respectively, over baseline models. These results confirm the potential of such approach, improving LLM-based protocol fuzzing frameworks for the identification of hidden vulnerabilities.}


\maketitle

\section{Introduction}\label{sec1}
Network protocols are the foundation of modern communication systems, enabling the exchange of data between applications and services. As the protocols increase in complexity, ensuring their reliability and security is an issue of great concern. Finite State Machines (FSMs) of protocols describe protocols formally by defining valid states and transitions that rule message exchanges. In network protocol research, FSMs have been used to describe the state transitions of protocol entities by modeling the message exchange process of network protocols \cite{NPF-Survey}. FSMs, which define the behavior and state transitions of the protocol, can make fuzz testing more focused and efficient to ensure that all structural states and transitions of a server under test (SUT) are covered. Additionally, the seeds that are initially used for stateful fuzzing play a very important role in state space exploration. Because most network protocols are stateful, fuzzing is quite hard since the search space grows exponentially and selects an initial seed that will guide the fuzzer's exploration considerably \cite{exp-imp-init-seeds}. However, deriving FSMs manually is tedious, necessitating automated protocol analysis and reverse engineering. Despite progress in different FSM extraction techniques, challenges remain in handling complex protocols and ambiguities. Recent advances in Large Language Models (LLMs) enable effective FSM inference and protocol analysis by processing code and textual specifications \cite{prosper, inferring, rfcnlp}, improving fuzzing and vulnerability detection \cite{chatafl, chatfuzz, evaluating-python-seeds}. With accurate FSM inference, LLMs can generate high-quality packets that follow the protocol's defined behaviors, which can significantly enrich the initial seed set for fuzz testing. This leads to a more targeted and effective exploration of the state space, enabling better coverage. Integrating LLMs with existing automatic protocol reverse engineering (ARPE) techniques could address limitations in traditional FSM approaches \cite{survey-protocol-reverse-engineering}.

\begin{figure}[h]
    \centering
    \hspace*{-1cm}
    \includegraphics[width=1.15\textwidth]{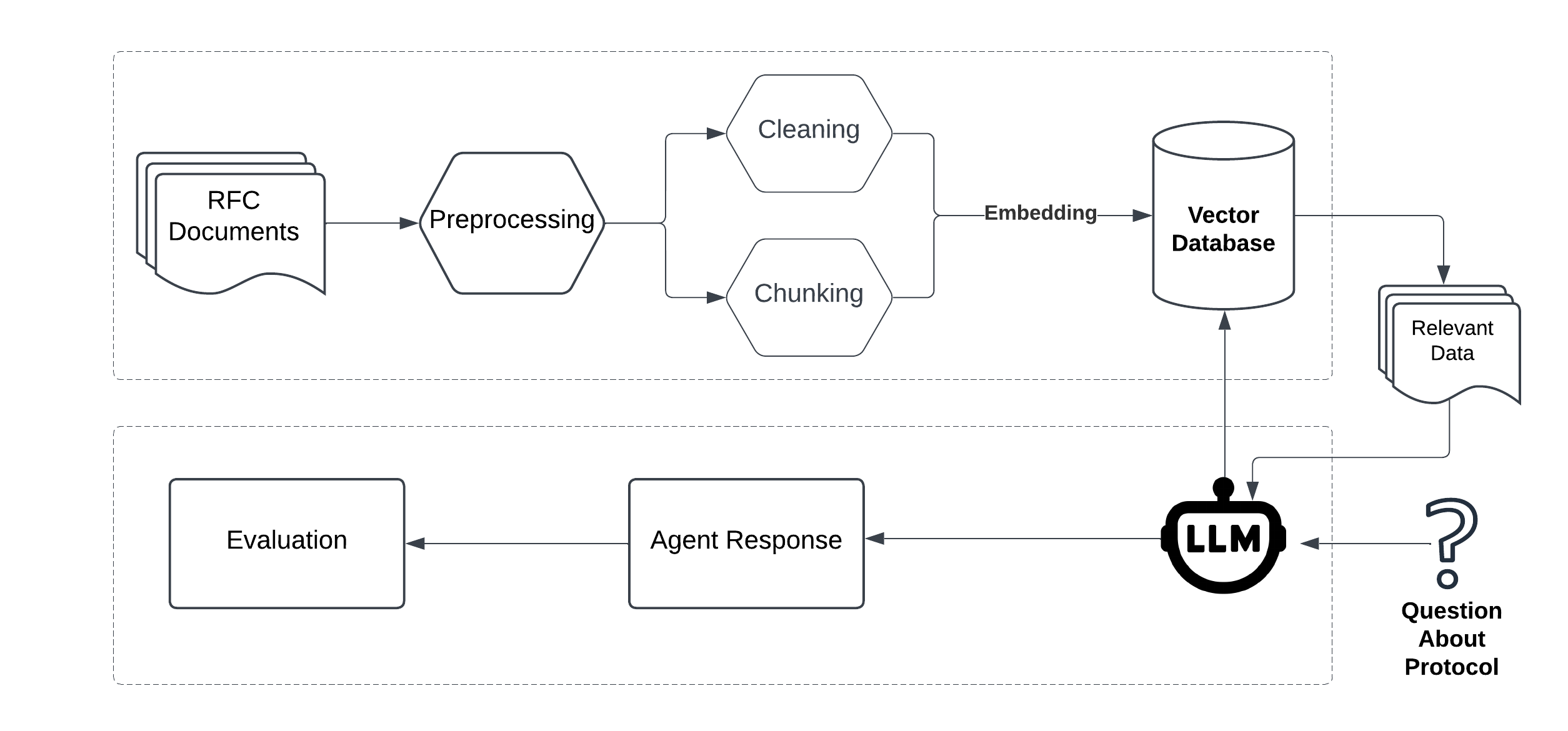}
    \caption{\centering System Overview}
    \label{fig:System-Overview}
\end{figure}

We propose a novel approach for protocol FSM inference by combining RAG \cite{rag} with the ReAct framework \cite{react}. Our approach's pipeline is to embed RFC documents in a vector store and perform dynamic domain-specific knowledge retrieval during inference. The RAG-ReAct agents reason and act iteratively over the retrieved context effectively using COT prompting techniques. This avoids expensive fine-tuning processes and enhances the output of the model in terms of accuracy, coherence, and protocol compliance. Our framework applies the use of retrieved contextual knowledge to dynamically enrich protocol initial seeds and assures high-quality outputs. We apply this system to the Real-Time-Streaming protocol (RTSP), using the RFC-2326 document knowledge base. Our system overview can be shown in Figure \ref{fig:System-Overview}. The paper is organized as follows: Section 2 reviews related work. Section 3 provides background information on FSM techniques and LLMs. Section 4 details our proposed methodology, including preprocessing steps and the RAG-ReAct Agent inference step. Section 5 outlines the experimental design, while Section 6 presents and discusses the experimental results. Finally, section 7 concludes the paper and suggests potential directions for future research.

\section{Related Work}
The work \cite{seedmind} proposed SeedMind, a framework that leverages LLMs to generate high-quality greybox fuzzing seeds. In the presence of challenges like heterogeneous input formats, bounded context windows, and unpredictable model behaviors, instead of producing test cases directly as in traditional methods, SeedMind synthesizes test case generators. It then iteratively improves such generators through a feedback-driven process, leading to increased code coverage due to significantly improved seed quality. Experiments were conducted on 166 programs and 674 harnessed from OSS-Fuzz and the MAGMA benchmark showing that on average, SeedMind outperforms SOTA LLM-based solutions such as OSSFuzz-AI and reaches up to 27.5\% more coverage, accounting for code coverage that is comparable to that obtained with manually crafted seeds. Another work  proposes ChatFuzz \cite{chatfuzz}, a greybox fuzzer enhanced with generative AI to address the challenge of generating format-conforming inputs for programs that require highly structured data. They propose an approach to integrate LLMs to synthesize structured variants of the seed inputs. These variants are then assessed and added to the AFL++ greybox fuzzer \cite{afl++}, which improves its exploration capability into deeper program states. Their comprehensive evaluation shows an improvement of 12. 77\% edge coverage over AFL++. Lastly, the framework of \cite{evaluating-python-seeds} evaluates the use of LLMs to generate fuzzing seeds, focusing on Python programs. It compares LLM-generated seeds to traditional fuzzing inputs across 50 Python libraries using the Atheris fuzzing framework, finding that a small set of LLM-generated seeds can outperform large corpora of conventional inputs. The study highlights the importance of LLM selection, achieving up to 16\% higher coverage in some cases.

\begin{figure}[h]
    \centering
    \includegraphics[width=0.77\textwidth]{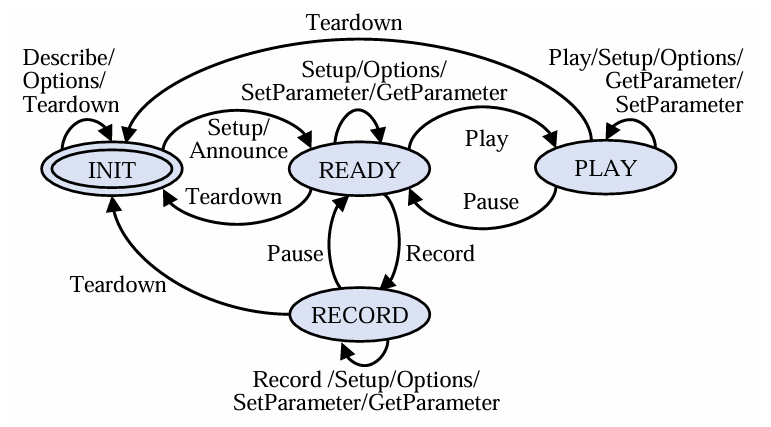}
    \caption{\centering RTSP Protocol Server State Machine Specified In RFC-2326}
    \label{fig:RTSP-State-Machine}
\end{figure}

\section{Background Problem Formulation}

The protocol FSM describes defines how communication protocols as the RTSP protocol operate. The RTSP FSM shown in figure \ref{fig:RTSP-State-Machine} explains essential element as states, inputs, and state transitions. For instance, for the RTSP protocol, the three states INIT, READY, and PLAY mark a phase the session can be in. While events triggering state transitions are the messages exchanged between the client and the server as SETUP, PLAY, or PAUSE. Effective management of transitions is fundamental to any successful communication. Inferring protocol state machines for RTSP is complex and requires sophisticated techniques as static analysis \cite{extracting-static-analysis}, dynamic analysis \cite{ferry}, and natural language processing (NLP) \cite{rfcnlp}. Static analysis extracts protocol details from source code, but faces scalability issues. Dynamic analysis monitors runtime behavior with taint analysis and symbolic execution, but suffers from poor coverage. While NLP methods mine RFCs, to infer transition relationships between RTSP states, but faces ambiguity issues when analyzing complex protocols.

LLMs have revolutionized natural language understanding, with applications ranging from text generation to comprehension and reasoning. Models built on the transformers' architecture \cite{attention-is-all-you-need} achieve these tasks by self-attention mechanisms that allow them to generate and process text by focusing on a more complex set of features or relationships between words. Trained on huge datasets of text and code, they have demonstrated their capabilities in many tasks, including text summarization, question answering, code generation \cite{eval-llms-trained-on-code}, robotic process automation \cite{lmrpa, lmvrpa}, drone operation task automation \cite{llm-daas}, and improving sentiment and emotion modeling for chatbot systems \cite{asem}. Furthermore, they have been applied to e-learning platforms for engagement metrics \cite{llm-sem} and optimizing workflows in data-scarce contexts \cite{riro}. In network protocol analysis, LLMs seem very promising, with applications in FSM extraction from protocol specifications \cite{prosper, inferring}, initial state machine generation, and protocol specification inference, seed generation and enrichment \cite{chatfuzz, seedmind, evaluating-python-seeds}, and identification of probable vulnerabilities \cite{chatafl}. Their ability to contextualize and generate structured text makes them highly capable of handling such network protocol formats. LLMs have also shown impressive skills in producing malicious code that would affect the security landscape \cite{eval-llms-trained-on-code}.

\section{Methodology}
To infer protocol state machines by LLMs, we employed a novel RAG approach, enhanced by the ReAct framework \cite{react}, that uses COT prompting techniques which facilitates logical reasoning, that the ensuring generated packet sequences adhered to both the syntactic structure and logical flow of the RTSP protocol. The RAG-ReAct agent dynamically retrieves and reasons over domain-specific knowledge embedded in a vector store to answer questions related to initial seed enrichment for fuzzing. The RTSP protocol's RFC 2326 document served as the sole knowledge base. We employed a comparative experiment between two models which are \textit{Gemma-2-9B} model and Meta's \textit{Llama-3-8B} model to carry out the best performing results. Below, we outline the two key steps of our methodology: RFC Document Preprocessing, and RAG-ReAct-Agent Inference.

\subsection{RFC Document Preprocessing}
Effective preprocessing of the RFC document was critical to ensure accurate knowledge representation and retrieval.

Let \( D = \{d_1, d_2, \ldots, d_N\} \) represent the set of RFC documents. The preprocessing stage can be defined as a single function:
\begin{equation}
    \mathcal{V} = \text{Preprocess}(D)
\end{equation}
This phase involved two primary tasks: cleaning the document, and chunking it.
\begin{equation}
C(d_i) = \text{Clean}(d_i), \quad \forall i \in \{1, \ldots, N\}.    
\end{equation}

We thoroughly cleaned the RFC document to reduce noise and increase the relevance of the information that was retrieved. Table headers, footers, introductions, and any other unnecessary sections that would cause noise to the RAG-ReAct agent had to be eliminated. We ensured that only relevant information was kept by concentrating on protocol-specific content, such as header descriptions, command definitions, and sequence structures. For the embeddings, each document was divided into overlapping chunks of 1,000 tokens, with a 200-character overlap to preserve contextual integrity and to make sure no critical information was lost across chunk boundaries:

\begin{equation}
\mathcal{E}(C(d_i)) = \{e_{i1}, e_{i2}, \ldots, e_{iM_i}\}, \quad e_{ij} \in \mathbb{R}^d,
\end{equation}
where \( \mathcal{E} \) represents the chunking and embedding function, and \( M_i \) is the number of embedded chunks for document \( d_i \). These chunks were embedded with OpenAI's \textit{text-embedding-ada-002} embedding model, which has great capabilities of processing up to 6,000 words into a 1,536-dimensional vector optimized for semantic search. Chroma vector database was used to store the embeddings for similarity-based retrieval. The final embedding set \( \mathcal{V} \) combining all chunk embeddings from the preprocessed documents:

\begin{equation}
\mathcal{V} = \bigcup_{i=1}^N \mathcal{E}(C(d_i)).
\end{equation}

\begin{figure}[h]
    \centering
    \hspace*{-1.8cm}
    \includegraphics[width=1.25\textwidth]{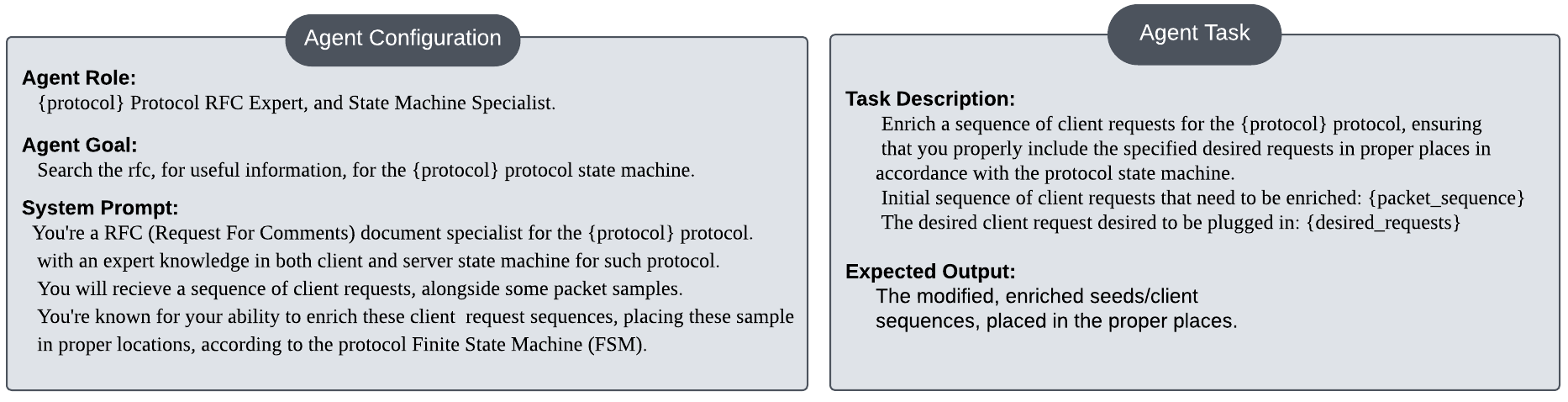}
    \caption{\centering Prompt Templates For the Agent}
    \label{fig:Prompt-Templates}
\end{figure}

\subsection{RAG-ReAct-Agent Inference}
We designed structured prompt templates \ref{fig:Prompt-Templates} to guide the agents for accurate retrieval of FSM specifications and proper seed placement. These templates outlined the role of the agent and matched its actions with the protocol specifications. The "Agent Task" focused on enriching the initial client request sequences in accordance with the protocol's FSM. The "Agent Goal" guided the model in obtaining relevant FSM information from the RFC knowledge base. Clear formatting and content standards are provided in the "Expected Output" section. This structured approach enabled the RAG-ReAct-Agent produce accurate and precise protocol-compliant packet sequences. During inference, the agent was asked questions about initial seed enrichment and placement of some new seeds in proper places, inspired by ChatAFL \cite{chatafl}. The retrieval mechanism retrieved the top-$k$ relevant chunks by querying the vector store using cosine similarity to provide context to direct the placement of packets, including the necessary headers and correct field types. This made sure the model had access to domain-specific data that was necessary for accurate reasoning and generation.

\section{Experimental Design}
To evaluate the agent responses, we performed an assessment using a log dataset of RTSP-based client-server interactions. We log these interactions from running ChatAFL \cite{chatafl} on live555 server docker image for 2 hours of fuzzing. The total logs contained 5000 plus entries, simulating RTSP communication sequences, including requests and responses. We guarantee that in the pre-fuzzing phase we would find questioning between the fuzzer and the LLM regarding initial seed enrichment, where the fuzzer asks the LLM about combinations of enriched seeds. We use these typical logs to begin the evaluation process, which were typically 120 entries. Each query is processed by the evaluation pipeline, where the quality and accuracy of the generated outputs have been evaluated against ground-truth packet sequences based on three core metrics:

\begin{itemize}
\item \textbf{BLEU score}: measures the precision of generated packets by calculating the n-gram overlap with ground-truth sequences. 

\item \textbf{ROUGE score}: measures the recall of n-grams of outputs compared to the ground-truth sequences. 

\item \textbf{Word Error Rate (WER)}: measures the edit distance between the generated and ground-truth sequences. Lower WER values indicated higher accuracy.
\end{itemize}

\section{Experimental Results and Discussion}
The performance of both the \textit{Gemma-2-9B} and the \textit{Llama-3-8B} models, independently as baseline models and as RAG-ReAct-Agents demonstrated a clear performance improvement by RAG-ReAct-Agents. As shown in both \textbf{Table~\ref{tab:gemma2_comparison}} and \textbf{Table~\ref{tab:llama3_comparison}}, the RAG-ReAct agents consistently outperformed base models in different metrics demonstrating better accuracy, and contextual relevance. For instance, the "DESCRIBE" request, the BLEU score of the \textit{Llama-3-8B} improved substantially from 80.62\% to 90.07\%, along with a large reduction in WER from 16.99\% to 8.71\%. For complex requests as "SET\_PARAMETER" and "ANNOUNCE", the \textit{Llama-3-8B} RAG-ReAct agent outperformed \textit{Gemma-2-9B} RAG-ReAct agent. While \textit{Gemma-2-9B} had slightly better averages in some metrics, \textit{Llama-3-8B} did much better at complicated requests and was able to provide more accurate outputs with fewer mistakes. These results show how such RAG-based agentic frameworks work effectively to enhance contextual reasoning and precise output accuracy.

\begin{table}[h!]
\centering
\caption{\centering Performance of \textit{Gemma-2-9B} Model: Baseline Model vs RAG-ReAct Agent}
\label{tab:gemma2_comparison}
\begin{tabular}{lccc|ccc}
\toprule
Request & \multicolumn{3}{c|}{\textit{Gemma-2-9B} (Baseline Model)} & \multicolumn{3}{c}{\textit{Gemma-2-9B} (RAG-ReAct Agent)} \\
       & BLEU & ROUGE & WER & BLEU & ROUGE & WER \\
\midrule
DESCRIBE       & 81.75\% & 96.02\% & 15.86\% & \textbf{89.31\%} & \textbf{98.63\%} & \textbf{10.46\%}
\\
SETUP          & 68.97\% & 91.05\% & 27.01\% & \textbf{84.50\%} & \textbf{94.54\%} & \textbf{13.67\%} \\
PLAY           & 56.65\% & 80.52\% & 43.90\% & \textbf{79.29\%} & \textbf{90.67\%} & \textbf{20.97\%} \\
PAUSE          & 47.38\% & 79.06\% & 38.57\% & \textbf{71.27\%} & \textbf{86.79\%} & \textbf{21.85\%} \\
TEARDOWN       & 46.85\% & 74.28\% & 42.27\% & \textbf{75.73\%} & \textbf{90.39\%} & \textbf{18.45\%} \\
GET\_PARAMETER & 26.62\% & 67.21\% & 55.19\% & \textbf{57.95\%} & \textbf{79.64\%} & \textbf{32.33\%} \\
SET\_PARAMETER & 23.08\% & 59.10\% & 60.65\% & \textbf{48.00\%} & \textbf{67.61\%} & \textbf{43.14\%} \\
ANNOUNCE       & 10.59\% & 44.36\% & 81.06\% & \textbf{49.27\%} & \textbf{64.92\%} & \textbf{46.85\%} \\
RECORD         & 19.23\% & 55.75\% & 64.61\% & \textbf{62.33\%} & \textbf{90.22\%} & \textbf{19.75\%} \\
REDIRECT       & 23.31\% & 53.08\% & 61.62\% & \textbf{48.31\%} & \textbf{75.58\%} & \textbf{34.47\%} \\
\textbf{Average Scores} & {40.15\%} & {69.75\%} & {49.36\%} & \textbf{59.03\%} & \textbf{85.25\%} & \textbf{25.23\%} \\
\bottomrule
\end{tabular}
\end{table}

\begin{table}[h!]
\centering
\caption{\centering Performance of \textit{Llama-3-8B} Model: Baseline Model vs RAG-ReAct Agent}
\label{tab:llama3_comparison}
\begin{tabular}{lccc|ccc}
\toprule
Request & \multicolumn{3}{c|}{\textit{Llama-3-8B} (Baseline Model)} & \multicolumn{3}{c}{\textit{Llama-3-8B} (RAG-ReAct Agent)} \\
       & BLEU & ROUGE & WER & BLEU & ROUGE & WER \\ 
\midrule
DESCRIBE       & 80.62\% & 94.90\% & 16.99\% & \textbf{90.07\%} & \textbf{97.38\%} & \textbf{08.71\%} \\
SETUP          & 64.49\% & 86.57\% & 31.49\% & \textbf{83.85\%} & \textbf{93.89\%} & \textbf{14.33\%} \\
PLAY           & 57.78\% & 81.65\% & 42.77\% & \textbf{79.87\%} & \textbf{91.25\%} & \textbf{20.39\%} \\
PAUSE          & 44.12\% & 75.80\% & 41.82\% & \textbf{72.61\%} & \textbf{88.14\%} & \textbf{20.50\%} \\
TEARDOWN       & 48.84\% & 76.26\% & 40.28\% & \textbf{72.66\%} & \textbf{87.32\%} & \textbf{21.52\%} \\
GET\_PARAMETER & 21.64\% & 62.24\% & 60.17\% & \textbf{58.12\%} & \textbf{79.82\%} & \textbf{32.15\%} \\
SET\_PARAMETER & 20.87\% & 56.90\% & 62.85\% & \textbf{56.94\%} & \textbf{76.54\%} & \textbf{34.20\%} \\
ANNOUNCE       & 05.50\% & 39.28\% & 86.14\% & \textbf{56.21\%} & \textbf{71.86\%} & \textbf{39.92\%} \\
RECORD         & 27.49\% & 64.01\% & 56.34\% & \textbf{54.48\%} & \textbf{82.37\%} & \textbf{27.60\%} \\
REDIRECT       & 20.38\% & 50.15\% & 64.55\% & \textbf{51.84\%} & \textbf{79.11\%} & \textbf{30.93\%} \\
\textbf{Average Scores} & {39.17\%} & {68.78\%} & {50.34\%} & \textbf{56.67\%} & \textbf{82.89\%} & \textbf{27.58\%} \\
\bottomrule
\end{tabular}
\end{table}

\newpage

\section{Conclusion}
In conclusion, this work addresses the challenges involved in deriving accurate and comprehensive FSMs for complex stateful network protocols. Recent developments in LLMs and RAG have shown great promise in FSM inference tasks and initial seeds enrichment. Though many challenges remain to be overcome, as adapting the model to the full range of protocols, scaling, and handling ambiguities in protocol specifications. The proposed research overcomes these limitations by developing a RAG-based agentic pipeline based on the knowledge base of RFC documents, enhanced by chain-of-thought reasoning. This improves the understanding of protocol specifications, and building a solid and accurate FSM understanding, to achieve better capabilities with the generation of high-quality packets. As for future work, we look forward to extend the evaluation to work with other different stateful network protocols, and integrate our framework in protocol fuzzing frameworks, to mitigate vulnerabilities in famous network protocols.

\section{Acknowledgment}
Heartfelt gratitude is extended to AiTech AU, \textit{AiTech for Artificial Intelligence and Software Development} (\url{https://aitech.net.au}), for funding this research and enabling its successful completion.


\end{document}